\pdfoutput=1

\documentclass[twocolumn,aps,prl,showpacs,superscriptaddress]{revtex4}

\usepackage{amsmath,amssymb}
\usepackage[colorlinks=true,citecolor=blue,urlcolor=blue]{hyperref}

\def\endproof{\vrule height6pt width6pt depth0pt}

\begin{document}


\title{The exclusivity principle forbids sets of correlations larger than the quantum set}


\author{Barbara Amaral}
 \email{barbaraamaral@gmail.com}
 \affiliation{Departamento de Matem\'atica, Universidade Federal de Minas Gerais,
 Caixa Postal 702, 30123-970, Belo Horizonte, MG, Brazil}
 \affiliation{Departamento de Matem\'atica, Universidade Federal de Ouro Preto,
 Ouro Preto, MG, Brazil}
 \affiliation{Universitat Aut\`onoma de Barcelona, E-08193 Bellaterra, Barcelona, Spain}

\author{Marcelo Terra Cunha}
 \email{tcunha@mat.ufmg.br}
 \affiliation{Departamento de Matem\'atica, Universidade Federal de Minas Gerais,
 Caixa Postal 702, 30123-970, Belo Horizonte, MG, Brazil}

\author{Ad\'an Cabello}
 \email{adan@us.es}
 \affiliation{Departamento de F\'{\i}sica Aplicada II, Universidad de
 Sevilla, E-41012 Sevilla, Spain}


\date{\today}



\begin{abstract}
We show that the exclusivity (E) principle singles out the set of quantum correlations associated to any exclusivity graph assuming the set of quantum correlations for the complementary graph. Moreover, we prove that, for self-complementary graphs, the E principle, {\em by itself} (i.e., without further assumptions), excludes any set of correlations strictly larger than the quantum set. Finally, we prove that, for vertex-transitive graphs, the E principle singles out the maximum value for the quantum correlations assuming only the quantum maximum for the complementary graph. This opens the door for testing the impossibility of higher-than-quantum correlations in experiments.
\end{abstract}



\maketitle




{\em Introduction.---}One of the most seductive scientific challenges in recent times is deriving quantum theory (QT) from first principles. The starting point is assuming general probabilistic theories allowing for correlations that are more general than those that arise in QT, and the goal is to find principles that pick out QT from this landscape of possible theories.

There are, at least, three different approaches to the problem. One consists of reconstructing QT as a purely operational probabilistic theory that follows from some sets of axioms \cite{Hardy01,Hardy11,DB11,MM11,CDP11}. A second approach consists of identifying principles that explain the set of quantum {\em nonlocal} correlations \cite{PR94,V05,PPKSWZ09,NW09,ABPS09,ABLPSV09, CSS10,GWAN11,FSABCLA12}. The third approach consists of identifying principles that explain the set of quantum {\em contextual} correlations \cite{Specker60,Bell66,KS67,KCBS08,Cabello08} without restrictions imposed by a specific experimental scenario \cite{Cabello13,SBBC13,CDLP12b,Yan13}.

This third approach is based on two observations: on one hand, that quantum contextual correlations, {\em i.e.}, quantum correlations for compatible (but not necessarily spacelike compatible) measurements provide a natural generalization of quantum nonlocal correlations that leaves room for a wider range of experimental scenarios, including systems that cannot be separated into parts or represented as tensor product of smaller spaces \cite{KCBS08,LLSLRWZ11,YO12,KBLGC12,ZUZAWDSDK13,AACB13} and for systems prepared in arbitrary quantum states \cite{Cabello08,BBCP09,KZGKGCBR09,ARBC09,YO12,KBLGC12,ZUZAWDSDK13,DHANBSC13}. The second observation comes from the graph approach to quantum correlations introduced in Ref.~\cite{CSW10}. 
To any experimental scenario one can associate an exclusivity graph ${\cal G}$ in which vertices represent all possible events in that experiment, {\em i.e.}, all the propositions of the type ``outcomes $a,\ldots,c$ occur for measurements $x,\ldots,z$'' and edges link mutually exclusive events.
By exclusive we mean that they correspond to exclusive outcomes of some common measurement.
Then the set of quantum correlations consists of all possible probability distributions allowed by QT for the vertices of ${\cal G}$.
Some important sets of probability distributions for the vertices of $\cal{G}$ can be defined.
The sets of probabilities that obey the hypothesis of noncontextuality of outcomes satisfy the so-called noncontextuality (NC) inequalities.
For a given experimental scenario, NC inequalities take the form
\begin{equation}\label{NCIneq}
S_w = \sum _i w_i p\left(v_i\right) \stackrel{\mbox{\tiny{NCHV}}}{\leq} \alpha\left(G,w\right),
\end{equation}
where $w_i \geq 0$ are weights for each corresponding vertex, $G \subset {\cal G}$ is the induced subgraph of all nonzero weight vertices in $S_w$, and $\alpha\left(G,w\right)$ is the weighted version of the independence number.
To any $S_w$ one can associate the weighted subgraph $(G,w)$, which is called the exclusivity graph of $S_w$. The fundamental observation \cite{CSW10} is that, in QT,
\begin{equation}\label{QIneq}
 S_w \stackrel{\mbox{\tiny{Q}}}{\leq} \vartheta\left(G,w\right),
\end{equation}
where $\vartheta(G,w)$ is the weighted Lov\'asz number of $G$ \cite{Lovasz79,GLS86}.

On the other way, for any given $G$, there is always some NC inequality experiment reaching $\vartheta\left(G\right)$ and spanning the set of probabilities allowed by QT for the vertices of $G$.
This set will be denoted ${\cal{Q}}\left(G\right)$. This shows that $\vartheta\left(G\right)$ is a fundamental physical limit for correlations
and ${\cal{Q}}\left(G\right)$ a fundamental physical set that appears when we remove any additional constraint imposed by a specific experimental scenario.
This suggests that a fundamental question for understanding quantum correlations is: {\emph{Which principle singles out this limit and this set}}?

The exclusivity principle (E) was proposed as a possible answer \cite{Cabello13}. It states that the sum of the probabilities of any set of pairwise mutually exclusive events cannot be higher than~1. The E principle is implied by Specker's observation that, in classical physics and QT, any set of pairwise compatible observables is jointly compatible \cite{Specker60,Specker09} and was first applied to general probabilistic theories by Wright \cite{Wright78}. In Ref.~\cite{CSW10} it was shown that, when we consider the experiment to test $S_w$ alone, then the maximum value of $S_w$ allowed by the E principle is given by the weighted version of the fractional packing of $G$, $\alpha^*(G,w)$ \cite{Shannon56}.
The principle of local orthogonality \cite{FSABCLA12} may be seen as the E principle restricted to Bell scenarios. However, when this restriction is removed is when the E principle shows itself more powerful, since while for a given graph $G$, there is always a NC inequality for which QT reaches $\vartheta\left(G\right)$ \cite{CSW10}, this is not true if ``NC inequality'' is replaced by ``Bell inequality'' \cite{SBBC13}.

For example, the E principle, applied to the exclusivity graph, singles out the maximum quantum value for some Bell and NC inequalities \cite{Cabello13}.
When applied to the OR product of two copies of the exclusivity graph (which may be seen as two copies of the same experiment), the E principle singles out
the maximum quantum value for experiments whose exclusivity graphs are vertex-transitive and self-complementary \cite{Cabello13},
which include the simplest NC inequality violated by QT, namely the Klyachko-Can-Binicio\u{g}lu-Shumovsky (KCBS) inequality \cite{KCBS08}.
Moreover, either applied to two copies of the exclusivity graph of the Clauser-Horne-Shimony-Holt (CHSH) \cite{CHSH69} Bell inequality \cite{FSABCLA12}
or of a simpler inequality \cite{Cabello13}, the E principle excludes Popescu-Rohrlich nonlocal boxes \cite{PR94} and provides an upper bound to the maximum violation of the CHSH inequality which is close to the Tsirelson bound \cite{Cirelson80}. In addition, when applied to the OR product of an infinite number of copies, there is strong evidence that the E principle singles out the maximum quantum violation of the NC inequalities whose exclusivity graph is
the complement of odd cycles on $n \ge 7$ vertices \cite{CDLP12b}. Indeed, it might be also the case that, when applied to an infinite number of copies, the E principle singles out the Tsirelson bound of the CHSH inequality \cite{FSABCLA12,Cabello13}.

Another evidence of the strength of the E principle was recently found by Yan \cite{Yan13}. By exploiting Lemma~1 in \cite{Lovasz79},
Yan has proven that, {\em if all correlations predicted by QT for an experiment with exclusivity graph $G$ are reachable in nature,} then the E principle singles out the {\em maximum} value of the correlations produced by an experiment whose exclusivity graph is the complement of $G$, denoted as $\overline{G}$.

In this Letter we shall prove three stronger consequences of the E principle. They show that the E principle goes beyond any other proposed principle towards the objective of singling out quantum correlations.


We will consider a sum of probabilities of events $\{e_i\}$, such that each event $e_i$ occurs with probability $P_i$ (\textit{i.e.}, all weights are chosen to be $1$ so hereafter we will write $G$ instead of $(G,w)$ and $S$ instead of $S_w$),
\begin{equation}
 S=\sum_{i} P_i.
\end{equation}
All these probabilities can be collected in a single vector $P$.


{\em Result 1:} Given the quantum set ${\cal Q}(\overline{G})$, the E principle singles out the quantum set ${\cal Q}(G)$.


{\em Proof:} Let $\{e_i\}$ be a set of $n$ events with exclusivity graph $G$ and $\{f_i\}$ be a set of $n$ events with exclusivity graph $\overline{G}$, such that $e_i$ and $f_i$ are independent.
Define the event $g_i$ which is true if and only if both $e_i$ and $f_i$ are true, $g_i = \left(e_i,f_i\right)$.
Note that the exclusivity graph of the events $\{g_i\}$ is the complete graph on $n$ vertices because $\{g_i\}$ is a set of pairwise mutually exclusive events.

Since $e_i$ and $f_i$ are independent $p(g_i)= P_i \bar{P}_i$, where $P_i=p\left(e_i\right)$ and $\bar{P}_i=p\left(f_i\right)$.
Using the E principle we have
\begin{equation}
 \label{eprinciple}
 \sum_i P_i \bar{P}_i \stackrel{\mbox{\tiny{E}}}{\leq} 1.
\end{equation}
Theorem 3.1 in Ref.~\cite{GLS86} implies that
\begin{equation}
 {\cal Q}(G)=\left\{P \in \mathbb{R}^n; P_v \geq 0, \vartheta(\overline{G},P) \leq 1\right\},
\end{equation}
where
\begin{equation}
\vartheta(\overline{G},P)= \max\left\{\sum_i P_i \bar{P}_i; \bar{P} \in {\cal Q}(\overline{G})\right\}.
\end{equation}
If the set of allowed distributions for $\overline{G}$ is ${\cal Q}(\overline{G})$, expression~\eqref{eprinciple} implies that the distributions in
$G$ allowed by the E principle belong to ${\cal Q}(G)$.\hfill\endproof



Physically, the proof above can be interpreted as follows: Assuming that nature allows all quantum distributions for $\overline{G}$, the E principle {\em singles out the quantum distributions for} $G$.

Result~1 does not imply that the E principle, by itself, singles out the quantum correlations for $G$, since we have assumed QT for $\overline{G}$.
Nonetheless, it is remarkable that the E principle connects the correlations of two, a priori, completely different experiments on two completely
different quantum systems. For example, if $G$ is the $n-$vertex cycle $C_n$ with $n$ odd, the tests of the maximum quantum violation of the corresponding
NC inequalities require systems of dimension $3$ \cite{CSW10,CDLP12b,LSW11,AQBTC13}. However, the tests of the maximum quantum violation of the NC inequalities with exclusivity graph $\overline{C_n}$, require systems of dimension that grows with $n$ \cite{CDLP12b}. Similarly, while two qubits are enough for a test of the maximum quantum violation of the CHSH inequality, the complementary test is a NC inequality that requires a system of, at least, dimension $5$ \cite{Cabello13d}.

An important consequence of Result 1 is that the larger the quantum set of $G$, 
the smaller the quantum set for $\overline{G}$, since each probability allowed for $G$ becomes
a restriction on the possible probabilities for $\overline{G}$. Such duality gets stronger when $G$
is a self-complementary graph.

A graph $G$ is self-complementary when $G$ and $\overline{G}$ are isomorphic. When $G$ is self-complementary, the graph $\overline{G}$ may be seen as another copy of the same experiment and,
as a corollary of Result~1, we have the following result:


{\em Result 2:} If $G$ is a self-complementary graph, the E principle, by itself, excludes any set of probability distributions strictly larger than the quantum set.

{\em Proof:} Let $X$ be a set of distributions containing ${\cal Q}(G)$ and let $P \in X \setminus{\cal Q}(G)$. By Result~1, 
there is at least one
$\bar{P} \in {\cal Q}\left(\overline{G}\right)$ such that
\begin{equation}
\sum_{i \in V(G)}P_i\bar{P}_i > 1,
\label{equationself}
\end{equation}
which is in contradiction with the E principle.
Since $G$ is self-complementary, after a permutation on the entries given by the isomorphism between $G$ and $\overline{G}$,
$\bar{P}$ becomes an element of ${\cal Q}(G)$ and hence $P$ and $\bar{P}$ belong to $X$.  Equation \eqref{equationself}
implies that this set is not allowed by the E principle.  \hfill\endproof

Notice that this proof applies to the KCBS inequality since its exclusivity graph is self-complementary. 



The exclusivity graphs of many interesting inequalities including CHSH \cite{CHSH69}, KCBS \cite{KCBS08}, the $n-$cycle inequalities \cite{CSW10,CDLP12b,LSW11,AQBTC13},
and the antihole inequalities \cite{CDLP12b} are vertex-transitive. A graph is vertex-transitive if for any pair $u,v \in V(G)$ there is $\phi \in {\rm Aut}(G)$ such that $v=\phi(u)$, where ${\rm Aut}(G)$ is the group of
automorphisms of $G$ (\textit{i.e.}, the permutations $\psi$ of the set of vertices such that $u, v \in V(G)$ are adjacent if and only if $\psi(u), \psi(v)$ are adjacent).


%

{\em Result 3:} If $G$ is a vertex-transitive graph on $n$ vertices, 
given the quantum maximum for $\overline{G}$, the E principle singles 
out the quantum maximum for $G$.


{\em Proof:} 
The proof comes in three steps. First we prove that
%
if $G$ is a vertex-transitive graph, then the quantum maximum for $S = \sum_i P_i$ is attained at the constant distribution $P_i = p_{\textrm{max}}$.

Let $P= \left(p(e_1),p(e_2),\ldots, p(e_n)\right)$ be a distribution reaching the maximum.
Given an automorphism of $G$, $\phi \in {\rm Aut}(G)$, consider the distribution $P_{\phi}$ defined as $p_{\phi}(e_i)= p(\phi(e_i))$.
This is also a valid quantum distribution, also reaching the maximum for $S$.
Define the distribution
\begin{equation}
 Q=\frac{1}{A} \sum_{\phi \in {\rm Aut}(G)} P_{\phi},
\end{equation}
where $A = \# {\rm Aut}(G)$.
Since $G$ is vertex-transitive, given any two vertices of $G$, $e_i$ and $e_j$, there is an automorphism $\psi$ such that $\psi(e_i)= e_j$. Then,
\begin{eqnarray}
q(e_j)&=& q(\psi(e_i)) \nonumber \\
 &=& \frac1A \sum_{\phi \in {\rm Aut}(G)} p_{\phi}(\psi(e_i)) \nonumber \\
 &=& \frac1A \sum_{\phi \in {\rm Aut}(G)} p\left( \phi \circ \psi(e_i)\right) \nonumber \\
 &=& \frac1A \sum_{\phi' \in {\rm Aut}(G)} p_{\phi'}(e_i) \nonumber \\
 &=& q(e_i).
\end{eqnarray}

The second step is to show that
if $G$ is a  vertex-transitive graph on $n$ vertices, then the E principle implies that the quantum maxima for $S(G)$ and for $S(\overline{G})$ obey 
\begin{equation}\label{eq:QuantumMaxima}
M_Q\!\left({G}\right) M_Q\!\left(\overline{G}\right) \stackrel{\mbox{\tiny{E}}}{\leq} n.
\end{equation}

This can be done using the above property for both, $G$ and $\overline{G}$. 
It implies $n p_{\textrm{max}} = M_Q\!\left(G\right)$ and $n \bar{p}_{\textrm{max}} = M_Q\!\left(\overline{G}\right)$.
Inequality \eqref{eprinciple} for these extremal distributions reads
\begin{equation}
n\,p_{\textrm{max}}\,\bar{p}_{\textrm{max}} \stackrel{\mbox{\tiny{E}}}{\leq} 1,
\end{equation}
which leads to Expression \eqref{eq:QuantumMaxima}.

The missing step is to prove that also 
$M_Q\!\left({G}\right) M_Q\!\left(\overline{G}\right) {\geq}\, n.$
This comes from the graph approach, which identifies the quantum maximum of $S$ with the Lov\'asz number of the corresponding exclusivity graph: 
\begin{subequations}
\begin{eqnarray}
\vartheta({G})&=&M_Q\left({G}\right),\\
\vartheta(\overline{G})&=&M_Q\left(\overline{G}\right),
\end{eqnarray}
\end{subequations}
and since
 for vertex-transitive graphs $\vartheta(G)\; \vartheta(\overline{G}) \geq n$ (Lemma~23 in Ref.~\cite{Knuth93}), the Result follows.
\hfill\endproof

Result~3  opens the door to experimentally discard higher-than-quantum correlations. 
Specifically, Inequality \eqref{eq:QuantumMaxima} implies that we can test that the maximum value
of correlations with exclusivity graph $G$ cannot go beyond its quantum maximum without violating the E principle, by performing an independent
experiment testing correlations with exclusivity graph $\overline{G}$ and experimentally reaching its quantum maximum \cite{Cabello13d}.
A violation of the quantum bound for $G$ in any laboratory would imply the impossibility of reaching the quantum maximum for $\overline{G}$ in any other laboratory.


{\em Conclusions.---}Here we have presented three results. First, we have shown that the E principle singles out the set of the quantum correlations associated to any exclusivity graph
assuming the set of quantum correlations for the complementary graph.
This result goes beyond the one presented by Yan in \cite{Yan13}, since using the same assumptions we have shown that the E principle singles out the
entire set of quantum correlations and not just its maximum.

Second, we have shown that the power of the E principle for singling out quantum correlations goes beyond
the power of any previously proposed principle. While previous principles cannot rule out the existence of sets of correlations strictly larger than the
quantum one for any single experiment \cite{PR94,V05,PPKSWZ09,NW09,ABPS09,ABLPSV09,CSS10,GWAN11,FSABCLA12}, we have shown that, for self-complementary graphs,
the E principle, by itself, excludes any set of correlations strictly larger than the quantum set.

Finally, we have shown that the E principle allows for experimental tests discarding higher-than-quantum correlations for
those correlations represented by vertex-transitive graphs. Interestingly, the CHSH Bell inequality is one of these cases.

All these results still do not prove that the E principle is {\em the} principle for quantum correlations.
However, what is clear at this point is that the E principle has a surprising and unprecedent power for explaining many puzzling predictions of quantum theory.


\begin{acknowledgments}
The authors thank A. J. L\'opez-Tarrida for his suggestions for improving this manuscript.
This work was supported by the Brazilian program Science without Borders, the Brazilian National Institute for Science and Technology of Quantum Information,
the Brazilian agencies Capes, CNPq, and Fapemig, and the Spanish Project No.\ FIS2011-29400 (MINECO).
\end{acknowledgments}

{\em Note added:} After submitting this paper, we have learned that A. B. Sainz {\em et al.} have also found Result~1 \cite{Sainz13}.



\end{document}